\documentstyle[12pt]{article}
\newcommand{\ku}{\left|\uparrow \right\rangle }
\newcommand{\kd}{\left|\downarrow \right\rangle }
\newcommand{\bu}{ \left\langle \uparrow \right| }
\newcommand{\bd}{ \left\langle \downarrow \right| }
\newcommand{\tr}{{\rm {\bf {t\!r}}}}
\begin{document}
\typeout{--- Title page start ---}

\thispagestyle{empty}
\renewcommand{\thefootnote}{\fnsymbol{footnote}}

\begin{tabbing}
\hskip 11.5 cm \= {Imperial/TP/92-93/03}
\\
\> {Fermilab-Pub-92/318-A}
\\
\> {hep-th/9309051}
\\
\> {To appear in PRD 1993}
\end{tabbing}

\vskip 1cm
\begin{center}
{\Large\bf    Following a ``collapsing'' wavefunction }
\vskip 1.2cm
{\large\bf Andreas Albrecht}\\
NASA/Fermilab Astrophysics Center\\
P.O.B. 500, Batavia, IL 60510\\
and\\
Blackett Laboratory\footnote{Permanent address since August 1992},
Imperial College, Prince Consort Road\\
London SW7 2BZ  U.K.
\end{center}
\vskip 1cm
\begin{center}
{\large\bf Abstract}
\end{center}

I study the quantum mechanics of a spin interacting with an
``apparatus''.  Although the evolution of the whole system is unitary,
the spin evolution is not. The system is chosen so that the spin
exhibits loss of quantum coherence, or ``wavefunction collapse'', of
the sort usually associated with a quantum measurement.  The system is
analyzed from the point of view of the spin density matrix (or
``Schmidt paths''), and also using the consistent histories approach.
These two points of view are contrasted with each other.
Connections between the results and the form of the Hamiltonian are
discussed in detail.

\vskip 1cm

\typeout{--- Main Text Start ---}

\section{Introduction}
\label{i}
A cosmologist must face the the issue of interpreting quantum
mechanics without the benefit of an outside classical observer.  By
definition, there is nothing ``outside'' the universe!  The
traditional role of an outside classical observer is to cause
``wavefunction collapse''.  This process causes  a definite outcome of
a quantum measurement to be realized, with the probability for a given
outcome determined by the initial wavefunction of the system being
measured.  It is common to view this process as something the can not be
described by a wavefunction evolving according to a Schr\"{o}dinger
equation, but which instead must be implemented ``by hand''.

There is a growing understanding
\cite{z73,z81,z82,z83,c&l83,wz86,u&z89,z91,h&p&z92,z92} that the essential
features of
wavefunction collapse {\em can} be present in systems whose evolution
is entirely unitary.  The key is the inclusion of an ``environment''
or ``apparatus'' within the Hilbert space being studied.  A {\em
sub}system
can then exhibit the non-unitary aspects of wavefunction collapse
even
though the system as a whole evolves unitarily.  The wavefunction can
then divide up into a number of different terms, each of which reflect
a different ``outcome''.  When there is negligible interference among
the different terms during subsequent evolution, the ``definiteness''
of the outcome is realized in a restricted sense: Each term evolves as
if the others were ``not there'', so a subsystem state within a given
term evolves with ``certainty'' that its corresponding outcome is the
only one.  None the less, the total wavefunction describes all possible
outcomes, and one is never singled out.

Some people object
to all the ``extra baggage'' or ``many worlds''\cite{e57} which result from
retaining all possible
outcomes.   However, this approach has the advantage of allowing
quantum mechanics to operate in a much more fundamental way,
and
{\em predict} which subsystems can play the role of classical
observers.
Unless the predictions are falsified, this approach can never be
shown to be wrong.

In what follows I present a simple toy system which is designed to
illustrate the essential features of a quantum measurement.
A very primitive ``apparatus'' is coupled to a two state
``spin".  The whole
spin-apparatus system evolves unitarily and remains in a
pure state, even as the two subsystems exhibit the
non-unitary evolution associated with the measurement.
Both the ``consistent histories'' point of view, and a more
conventional point of view (using reduced density matrices
or ``Schmidt paths") are use to analyze the same process.
The connections between the two points of view are discussed.

This paper is an expanded version of  a talk presented at the
``Workshop on time asymmetry" in Mazagon Spain \cite{a91t}.
The results are the same, but in this paper I describe the
Hamiltonian, and explore in detail the relationship between
the  results and my choice of Hamiltonian.  The purpose is to
develop some intuition as to what attributes make a ``good
classical observer''.  I also discuss the relationship between the
``consistent histories'' and more traditional approaches in more detail.

The organization of this paper is as follow:  Section 2 presents
some mathematical tools.  Section 3 introduces the toy model and
illustrates what it has to do with a quantum measurement by
analogy with the double slit experiment.  Section 4 shows the
behavior of the toy model in more quantitative detail.
Section 5 introduces the ``consistent histories'' point of view,
and section 6 applies this point of view to the toy model.
Section 7 explains specifically
how the form of the Hamiltonian allows the density matrix
evolution described in section 4 to be achieved. Section
8 explains how the form of the Hamiltonian allows the
consistent histories described in Section 6 to be achieved.  Section
9 discusses the relationship between the Schmidt paths and consistent
histories.
Section 10  explores the fundamental role played by the
statistical ``arrow of time'' in the processes under study.
Conclusions are presented in Section 11, and a number of
technical results are presented in the Appendices.
Units in which $\hbar = 1$ are used throughout.

\section{Tools}
\label{t}

The focus of this paper is the evolution of the spin and apparatus
subsystems from pure into mixed states due to correlations being set
up between the subsystems.  The ``Schmidt decomposition"
provides a useful tool for dealing with these issues, and
it will be used throughout this paper.

If a closed system in a pure state $|\psi\rangle$ is divided into two
subsystems, one might want to think about $|\psi\rangle$'s of the form
\begin{equation}
|\psi\rangle = |\psi\rangle_1 \otimes |\psi\rangle_2.
\label{dpf}
\end{equation}
One could then say subsystem 1 is in the pure state $|\psi\rangle_1$,
and
subsystem 2 is in the pure state  $|\psi\rangle_2$.  However, this ``direct
product'' form for $|\psi\rangle$ is far from general.  A general
$|\psi\rangle$ would look like:
\begin{equation}
|\psi\rangle = \sum_{i,j} \alpha_{ij}|i\rangle_1 \otimes|j\rangle_2
\label{genform}
\end{equation}
where $\{|i\rangle_1\}$ and $\{|j\rangle_2\}$ span the two respective
subspaces.  Then one can not talk about  pure states for subsystems
1 and 2.   However, one can say something along these lines if
$|\psi\rangle$ takes the special form:
\begin{equation}
|\psi\rangle = \sum_{i} \alpha_{i}|i\rangle_1 \otimes|i\rangle_2
\label{sform}
\end{equation}
for some orthonormal sets $\{|i\rangle_1\}$ and $\{|i\rangle_2\}$.
This form is special because there is only one summation, and each
state for system 1 is uniquely correlated with a specific state for
system 2.
The reduced density matrix of system 2 is then
\begin{equation}
\rho_2 \equiv \tr_1\left(|\psi\rangle \langle\psi|\right) =
\sum_i\alpha^{\ast}_i\alpha_i|i\rangle_2{ }_2\!\langle i|
\label{rho2}
\end{equation}
which is {\em diagonal} in the $\{|i\rangle_2\}$ basis.  The result is
that the probability assigned to any state $|x\rangle_2$ of the spin is
\begin{equation}
\langle x|\rho_2|x\rangle = \sum_i \alpha^{\ast}_i\alpha_i
\left|\langle x| i\rangle_2\right|^2,
\label{prob2}
\end{equation}
which is an incoherent sum over the probabilities of each state
$|i\rangle_2$, weighted by the probability $ \alpha^{\ast}_i\alpha_i$
assigned to that particular state.

One can thus regard
system 2 to be in state $|i\rangle_2$ with probability
$\alpha^{\ast}_i\alpha_i$.  Although quantum mechanics allows one to
assign probabilities for the spin to be in any state, the basis in
which $\rho_2$ is diagonal is special, because only in that basis does
any matrix element of $\rho_2$ take the form of  an {\em incoherent}
sum as depicted in Eq (\ref{prob2})  (with no interference terms like
$\langle x | i \rangle_2\; _2\!\langle j | x \rangle$) .

It turn out that the ``special form'' of Eq (\ref{sform}) can always
be realized.  It is called ``Schmidt'' form, and follows directly from the
fact that any density
matrix can be diagonalized.  The Schmidt bases, $\{|i\rangle_1^S\}$
and $\{|i\rangle_2^S \}$, are the eigenstates of the reduced density
matrices $\rho_1$ and $\rho_2$, and $\alpha_{i} = \sqrt{p_i}$, where
$p_i$ are the eigenvalues ($\rho_1$ and $\rho_2$ have the same
eigenvalues, and the larger one has additional zero eigenvalues).
For more discussion of the Schmidt decomposition see
\cite{s07,s35,z73,k&z73,z90,b&k90,a92q}.  Ref \cite{s07} contains the original
mathematical
result, and a brief proof is offered in \cite{a92q}.

The Schmidt decomposition allows one to expose precisely
which correlations are present between two subsystems.  The
special form of Eq \ref{sform} shows that state $|1\rangle_1$
is uniquely correlated with state $|1\rangle_2$ and so on.
It also allows one to make the clearest possible statement of
the ``state of a subsystem", by providing the eigenstates and
eigenvalues of the relevant reduced density matrix.

\section{The toy model}
\label{ttm}

\subsection{Defining the model}
\label{dtm}

The toy model discussed here is two state ``spin''(system 2) coupled to a
modest sized ``apparatus'' (system 1).
The Hamiltonian
is the same one used in \cite{a92q},  which takes the form
\begin{equation}
H = H_1 \otimes I_2 + I_1 \otimes H_2  + H_I
\label{hform}
\end{equation}
where $I_k$ represents the unit operator in the space of subsystem $k$.
The first two terms give self Hamiltonians of the apparatus and spin
respectively, and the last term gives the
interaction between spin and apparatus.

For this article I have chosen the  parameters so that the interaction
Hamiltonian
dominates over the  self-Hamiltonians of the two subsystems.
(Specifically $E_1 = E_2 = 0.1$
and $E_I = 10$ in the notation of \cite{a92q}.) The size
of the system 1 is $n_1 = 25$ here, as opposed to $n_1 = 12$ in \cite{a92q}.

The interaction Hamiltonian is:
\begin{equation}
H_I = E_I\left( \ku \bu \otimes H^{\uparrow}_1 \; +
\; \kd\bd \otimes H^{\downarrow}_1\right)
\label{iham}
\end{equation}
where $H^{\uparrow}_1$ and $H^{\downarrow}_1$
are two {\em different} random Hermitian matrices in the system 1
subspace.  (Each independent real and imaginary part of each element
of $H^{\uparrow}_1$ and $H^{\downarrow}_1$ is chosen randomly on the
interval $[-0.5,0.5)$.)  The random matrices are chosen once and for
all at the start of the calculation, so $H_I$ is time independent.
For our purposes, the role of the self-Hamiltonians for the subsystems
can be described by the statement ``the deviation of
the total Hamiltonian from $H_I$ is very small''.
For more details see \cite{a92q} (A similar model is used in \cite{t92}).

The idea behind the form of $H_I $ is very simple:  If the spin is up
the apparatus is pushed in one direction by $H^{\uparrow}_1$ and
if the spin is down,  the apparatus is pushed in a very different
direction by $H^{\downarrow}_1$.  The goal is to correlate different
states in the apparatus with the $\ku$ and $\kd$ states for the spin.

\subsection{The purpose of the model}
\label{tpotm}

The toy model is designed to perform a very specific function:
The model should take an initial state of
the form
\begin{equation}
|\psi_i\rangle = \left(a\ku_2 + b\kd_2\right) \otimes |X\rangle_1
\label{ic}
\end{equation}
and evolve it into the state:
\begin{equation}
|\psi_f\rangle = a\ku_2\otimes|Y\rangle_1 + b\kd_2\otimes |Z\rangle_1
\label{fc}
\end{equation}
Where $\langle Y|Z\rangle = 0$.
Actual numerical results showing this evolution are presented in
Section \ref{r}, and a detailed analysis of why this model is able to
achieve these results appears in Sections \ref{teatdm}.

Both Eqs (\ref{ic}) and (\ref{fc}) are
in Schmidt form.  Thus one can see that initially the spin (and the
apparatus) is in a ``pure'' state.  Later, $\rho_2$ develops {\em
two} non-zero eigenvalues, so the spin is in a mixed state.  The eigenstates
of the final spin density matrix are $\ku$ and $\kd$.  Thus, at the
end the spin is clearly no longer in the $a\ku_2 + b\kd_2$ state, but
it may be said to be in an incoherent superposition of $\ku$ and $\kd$.  An
important feature is that the probabilities  assigned to $\ku$ and
$\ku$ at the end ($a^{\ast}a$ and $b^{\ast}b$ respectively) are the
 same as those assigned initially. The only difference is that
the initial state is a {\em coherent} superposition of $\ku$ and $\kd$.
The choice of $\ku$ and $\kd$ as the final eigenstates of
the density matrix was built into the dynamics (and the choice of
initial state).  Although the evolution of the spin is non-unitary
(since the eigenvalues of its density matrix change), the evolution of the
total spin plus apparatus system is chosen to be completely unitary.

\subsection{Analogy with the double slit}
\label{awtds}

What does this have to do with wavefunction collapse?  One might
expect  a parallel description of the standard
double slit experiment:  After passing through a
double slit, an electron wave packet becomes spread out into a
distinctive double slit diffraction pattern. At this point the
electron is still in a pure state, and it is at this point that
I wish to make
the analogy with Eq \ref{ic}, the initial state for the toy
system. After interacting with  a screen, the electron is
certainly not in a pure state, but
the electron may be expressed as an incoherent superposition of
localized packets. The probability assigned to each packet is the same
probability assigned to that location by the original pure electron
state.  (Extremely low probabilities are assigned at nodes of the double slit
pattern, for example.)  The loss of coherence of the initial state is due
to the
setting up of correlations between the electron and the screen.  The
screen plays the role of system 1 in equation (\ref{fc}) (of
course there would be more than two terms in the Schmidt
expansion of the electron-screen system).

For each localized packet the
screen is in a different orthogonal state. The extent to which the
electron density matrix eigenstates tend to be localized packets
rather than some other types of states is determined by the nature of
the interaction, and the initial state of the screen.  It is natural
to expect the eigenstates to be localized, due to the local nature of
interactions.  Because of the correlation between the screen and the
electron, one can determine the state of the electron by measuring the
state of the screen.  In fact, one normally does look at the screen,
not at the electron.

There are three key feature of  the double slit
experiment which are present in the toy system.  First, the density
matrix eigenstates (or Schmidt states) take a particular form after
the measurement
which is determined by the interactions.  This ``pointer basis''
(\cite{z82}) is $\{ \ku,\kd\}$ in the toy model, and fairly
localized wave packet states for the electron in the double slit example.

Second, the probability assigned to each density matrix eigenstate
after the measurement  corresponds to the same probability assigned
to that state in the pre-interaction pure state.  For the spin, this
results  because the coefficients ``$a$'' and ``$b$'' are the same
in Eqns \ref{ic} and \ref{fc}.   For the double slit case, the
diffraction pattern is represented in the distribution of density
matrix eigenvalues after the interaction.
(This is why, after many electrons strike the screen,
the diffraction pattern is produced.)

Third, it is very unlikely that the process will reverse itself.
For the
double slit, it is extremely unlikely that the screen will emit
an electron in a double slit diffraction pattern.   The reason
why the toy model is unlikely to evolve from Eq \ref{fc} back
to Eq \ref{ic} will be discussed in Section \ref{teatdm}.

One way the analogy does not work is in the details of the apparatus.
The apparatus in the toy model is much less sophisticated than  a
realistic screen.  Although the different ``outcomes'' of the
measurement do get correlated with orthogonal states of
the apparatus, the apparatus states do not represent a
nice ``pointer'' or ``mark on a screen'' which clearly
reflects the state of the quantity being measured.

\section{Results}
\label{r}

Figure 1 shows information about the spin as the whole system evolves.
Initially, the state is given by Eq (\ref{ic}), with $a = 0.7, b =
0.3$.  In the lower plot, the solid curve gives  $p_1$,
the largest eigenvalue
of $\rho_2$.  It starts out at unity, as required by the ``pure
state'' form of the initial conditions, and evolves to $0.7$, where it
holds steady.  The dashed curve gives the entropy, $S$,  of the spin ($S
\equiv - \tr[\rho_2 log_2(\rho_2)]$), in units where the maximum possible
entropy in unity.  The entropy starts out zero and increases.  This is always
the case when a system evolves from a pure to a mixed state.  (Note
the the combined ``spin $\otimes$ apparatus'' system remains in a
pure state, so {\it its} entropy is zero)

In the upper plot,  the dashed curve gives the overall probability for the
spin to be up, given by $\bu\rho_2\ku$.  This quantity is a ``constant
of the motion''.  The solid curve gives $|\langle \uparrow|
1\rangle^S|^2$, where $|1\rangle^S$ is the eigenstate of $\rho_2$ (or
``Schmidt state'')
corresponding to the largest eigenvalue.

Since $|1\rangle^S$ belong to
a two state Hilbert space, it is completely specified by $|\langle
\uparrow|1\rangle^S|^2$, up to an overall phase.  One can see that as
the eigenvalue ($p_1$) approaches $0.7$, the eigenvector becomes
essentially $\ku$.  Thus the behavior promised in the previous
section
(Eqs (\ref{ic}) and (\ref{fc})) is realized to a good accuracy.

Figure 2 is another representation of the way the eigenstates of
$\rho_2$ evolve.  The first row represents $|1\rangle^S$, and the second
row represents the other eigenvector.  The three columns
correspond to three times.  The histogram in each plot provides two
numbers, $p({\uparrow}) \equiv  |\langle \uparrow|
1\rangle|^2$ and $p({\downarrow}) \equiv |\langle \downarrow |
1\rangle|^2$ for the first row, and similarly for the second
eigenvector in the second row. In this way one can visualize a
``collapsing wavefunction''
by following the eigenstates of $\rho_2$ as they ``collapse'' onto the
$\{\ku, \kd\}$ basis.

\section{Consistent Histories}
\label{ch}

I will now make contact with the
``consistent histories'' or ``decoherence functional'' approach to
quantum mechanics of closed systems.  Until now I have been using the
wavefunction to assign instantaneous probabilities to different states
over a range
of times.  By contrast, the consistent histories endeavors to assign
probabilities to {\em histories}.
Consider two projection operators:
\begin{equation}
\hat{P}_{\uparrow} \equiv |{\uparrow}\rangle\!\langle {\uparrow}|
\otimes I_1; \; \; \;
\hat{P}_{\downarrow} \equiv |{\downarrow}\rangle\!\langle
{\downarrow}| \otimes I_1
\label{pdef}
\end{equation}
where $I_1$ is the identity operator in the apparatus subspace, and
$\{ |{\uparrow}\rangle, |{\downarrow}\rangle\}$ form an orthonormal
``projection basis''
which spans the
spin subspace.  These projection operators sum to unity:
\begin{equation}
\hat{P}_{\uparrow} + \hat{P}_{\downarrow} = I.
\label{sump}
\end{equation}
One can take the formal
expression for the time evolution:
\begin{equation}
|\psi (t)\rangle = e^{-iHt}|\psi(0)\rangle
\label{te}
\end{equation}
and insert the unit operator ($\hat{P}_{\uparrow} +
\hat{P}_{\downarrow}$) at will,
resulting, for example, in the identity:
\begin{eqnarray}
|\psi (t)\rangle &  = & (\hat{P}_{\uparrow} +
\hat{P}_{\downarrow})e^{-iH(t - t_1)}(\hat{P}_{\uparrow} +
\hat{P}_{\downarrow})e^{-iHt_1}|\psi(0)\rangle \\
& = & \hat{P}_{\uparrow}e^{-iH(t -
t_1)}\hat{P}_{\uparrow}e^{-iHt_1}|\psi(0)\rangle
+  \hat{P}_{\uparrow}e^{-iH(t -
t_1)}\hat{P}_{\downarrow}e^{-iHt_1}|\psi(0)\rangle \nonumber \\
& & +  \hat{P}_{\downarrow}e^{-iH(t -
t_1)}\hat{P}_{\uparrow}e^{-iHt_1}|\psi(0)\rangle
+  \hat{P}_{\downarrow}e^{-iH(t -
t_1)}\hat{P}_{\downarrow}e^{-iHt_1}|\psi(0)\rangle \\
& \equiv & |[{\uparrow},{\uparrow}]\rangle +
|[{\uparrow},{\downarrow}]\rangle +
|[{\downarrow},{\uparrow}]\rangle + |[{\downarrow},{\downarrow}]\rangle.
\label{pid1}
\end{eqnarray}
The last line just defines (term by term) a shorthand notation for the
previous line.  Each term represents a particular choice of
projection at each time, and in that sense corresponds to a particular
``path''.  In the path integral formulation of quantum mechanics
the time between projections is taken arbitrarily small, and the
time evolution is viewed as a sum over
microscopic paths.  For present purposes,
the time intervals can remain finite, representing a ``coarse
graining'' in time.  Each term in the above expression is called a ``path
projected state'', and the sum is a sum over coarse grained paths.

One attempts to assign the probability ``$\langle
[i,j]|[i,j]\rangle$'' to the path $[i,j]$, but to make sense, the
probabilities must obey certain sum rules.  For example, one can define
\begin{equation}
|[{\uparrow},\cdot]\rangle \equiv |[{\uparrow},{\uparrow}]\rangle +
|[{\uparrow},{\downarrow}]\rangle,
\label{oproj}
\end{equation}
where the ``$\cdot$'' signifies that {\it no} projection is made at
$t_1$.  One would want the probability for the
path $[{\uparrow},\cdot]$ to be
the sum of the probabilities of the two paths of which it is composed:
\begin{equation}
\langle [{\uparrow},\cdot] | [{\uparrow},\cdot]\rangle = \langle
[{\uparrow},{\uparrow}] | [{\uparrow},{\uparrow}]\rangle
+ \langle [{\uparrow},{\downarrow}] | [{\uparrow},{\downarrow}]\rangle
\label{sumrule}
\end{equation}
However,  one can ``square'' Eq (\ref{oproj}) to give the general
result:
\begin{equation}
\langle [{\uparrow},\cdot] | [{\uparrow},\cdot]\rangle = \langle
[{\uparrow},{\uparrow}] | [{\uparrow},{\uparrow}]\rangle
+ \langle [{\uparrow},{\downarrow}] | [{\uparrow},{\downarrow}]\rangle
+ \langle [{\uparrow},{\uparrow}] | [{\uparrow},{\downarrow}]\rangle
+ \langle [{\uparrow},{\downarrow}] | [{\uparrow},{\uparrow}]\rangle
\label{sumid}
\end{equation}
Only if the last two terms (the cross-terms) in Eq (\ref{sumid}) are
small is the sum
rule (Eq (\ref{sumrule})) obeyed.  When the relevant sum rules are
obeyed the paths are said to give ``consistent'' or ``decohering''
histories.  Advocates of this point of view argue that the only
objects in quantum mechanics which make physical sense
are sets of consistent histories.
For a discussion of how this simple example links up with
the (much more general) original work on this subject (
\cite{g84,o88a,o88b,o88c,g-m&h90})
see \cite{a92q}.  Other work on the consistent histories approach
includes \cite{jh90,o90a,g-m&h91,d&h92,o92}.

\section{Testing for consistent histories}
\label{tfch}

Table 1a checks the probability sum rule (Eq
(\ref{sumrule})) for the toy model whose evolution is depicted in Fig
1.  The projection times are $t_1 = .15 , t = .2 $, and the projection
basis is $\{\ku, \kd\}$.  The sum rule is obeyed to the accuracy
shown.  In fact, using the $\{\ku, \kd\}$ projection basis, the sum
rule is obeyed no matter which projection times are chosen and
how frequently the projections are made.
\begin{table}[t]
\begin{tabular}{cc}
\mbox{\begin{tabular}{c|c}
\multicolumn{2}{c}{Table 1a} \\
path  & value  \\
\hline
&   \\
$ \langle [ \uparrow \uparrow ] | [\uparrow \uparrow] \rangle  $ &
$0.70$
\\
 &  \\
$ \langle [ \uparrow \downarrow ] | [\uparrow \downarrow] \rangle  $ &
$0.00$
\\
 &  \\
$ \langle [ \uparrow \cdot ] | [\uparrow \cdot] \rangle  $ &
$0.70$
\\
 &  \\
\hline
 &  \\
\% violation & 0\%  \\
 &  \\
\hline
\end{tabular}
} &
\mbox{\begin{tabular}{c|c}
\multicolumn{2}{c}{Table 1b} \\
path  & value  \\
\hline
&   \\
$ \langle [ {\cal I} {\cal I} ] | [{\cal I} {\cal I}] \rangle  $ &
$0.74$
\\
 &  \\
$ \langle [ {\cal I} \perp ] | [{\cal I} \perp] \rangle  $ &
$0.03$
\\
 &  \\
$ \langle [ {\cal I} \cdot ] | [{\cal I} \cdot] \rangle  $ &
$0.61$
\\
 &  \\
\hline
 &  \\
\% violation & 25\%  \\
 &  \\
\hline
\end{tabular}
}
\end{tabular}
\caption{Testing the probability sum rule (Eq (15)) for different
paths.  For 1a the sum
rules are obeyed for any choice of $t_1$ and $t$. For 1b, $t_1 = .035$
and $t=0.06$ }
\label{table1}
\end{table}

This result came as a surprise to me.  After all the interesting
dynamics described in Figs 1 and 2, the consistent
histories approach offers a completely {\em static} perspective. The
constant  ``$\uparrow$'' and ``$\downarrow$'' paths are consistent
right through the period when the correlations are being set up.

One of the very interesting features of the consistent histories point
of view is that typically there are many different sets of consistent
histories.
It turns out that
for this particular example some sets of consistent histories
reflect the ``quantum measurement" more directly.

Consider for a moment a static (Hamiltonian $ = 0$) spin, not coupled
to any apparatus.
It turns out that as long as the same projection basis is chosen at
$t$ and $t_1$, one always gets consistent histories.  This is true for
any projection basis.  One could choose
$\{\ku, \kd\}$ or one could choose the projection
basis $\{|{\cal I}\rangle, |\perp\rangle\}$, where $|{\cal I}\rangle$ is
the initial
state of the spin ($a\ku_2 + b\kd_2$), and $|\perp\rangle$
is the state orthogonal to it.
A static spin would naturally result in  unit probability for the
$[{\cal I}, {\cal I}]$ path, and zero probability for all other paths.
Table 1b shows the results for the fully interacting spin, using
the  {$\{|{\cal I}\rangle, |\perp\rangle\}$} projection basis, but
otherwise the same as Table 1a.  Clearly the sum rules are not
obeyed in this case.

When $\{\ku, \kd\}$ was used as a projection basis, there was
no difference, from the consistent histories point
of view, whether the interactions between spin and
apparatus were present or not.  Consistent histories resulted in
either case.   When the $\{|{\cal
I}\rangle, |\perp\rangle\}$
projection basis was used, the effects of the interactions were
evident:  Only in the absence of interactions were those histories
consistent.

Gell-Mann and Hartle \cite{g-m&h90,g-m&h91} have emphasized the
important role that ``records" or correlations among subsystems can
have in producing consistent histories.   In \cite{a92q} I noted
that since the Schmidt decomposition gives an exact account
of whatever correlations are present, the Schmidt states
(eigenstates of the reduced density matrix) often make a very
good choice of projection basis.

Indeed, I have found the following types of histories are
always consistent for this toy system:  For the first projection
time one chooses the eigenstates of $\rho_2$ (the Schmidt states)
as the projection basis.  At the second projection time one
chooses the
$\{ \ku, \kd\}$ projection basis.  These paths are consistent
for any choices of the two projection times.  These paths certainly reflect the
measurement process, since (as shown in Figs 1 and 2) this process
shows up quite explicitly in the behavior of the Schmidt states.
One can expand on this set of paths by including additional
projections on the $\{ \ku, \kd\}$ basis.
However, one will not get
consistent histories if one makes additional projections on the Schmidt
basis (until after it coincides with the $\{ \ku, \kd\}$ basis).  Thus
the actual picture presented by any of these sets of paths is quite
different from the Schmidt paths depicted in Fig 1.

\section{Time evolution and the density matrix}
\label{teatdm}

This section, and the one which follows, are devoted to
describing how the Hamiltonian which governs
the toy system is related to the results presented above.

To explore the effect of $H_I$, note that the initial state
(Eq \ref{ic}) can be written:
\begin{equation}
|\psi_i\rangle =
a\ku\otimes |X\rangle_1\;+\;b\kd\otimes |X\rangle_1.
\label{icex}
\end{equation}
Under time evolution according to  $H_I$,  this state maintains
a similar form:
\begin{equation}
|\psi(t)\rangle =
a\ku\otimes |X_{\uparrow}(t)\rangle_1\;+\;b\kd\otimes
|X_{\downarrow}(t)\rangle_1.
\label{fcex}
\end{equation}
Where $|X_{\uparrow}(t)\rangle_1$ and $|X_{\downarrow}(t)\rangle_1$ are
the initial apparatus state $|X\rangle$ evolved forward in
time according to $H^{\uparrow}_1$ and $H^{\downarrow}_1$ respectively.
Because $H^{\uparrow}_1$ and $H^{\downarrow}_1$ are different
randomly chosen operators, on average the states $|X_{\uparrow}
(t)\rangle_1$ and $|X_{\downarrow}(t)\rangle_1$
have no more overlap than two randomly chosen vectors in the
apparatus subspace.  For sufficiently large apparatus subspaces,
the overlap will be extremely small.

{}From this point of
view the initial conditions,
where $\langle X_{\uparrow}(0)|X_{\downarrow}(0)\rangle = 1$, are
very special.  As time evolves the value of $\langle
X_{\uparrow}(0)|X_{\downarrow}(0)\rangle $ decreases to
its more natural small value. If $\langle X_{\uparrow}(0)|
X_{\downarrow}(0)\rangle $ were to become close to unity
later in time, this would correspond to the apparatus
``forgetting'' its records, analogous to the screen re-emitting an
electron in a diffraction pattern state.  In the toy model,
this happens very rarely (for large apparatus) because
two vectors evolving randomly in a large space rarely
overlap.  (This effect has been discussed at length
in this context by Zurek \cite{z82}.)

Much of the earlier discussion has focused on the eigenstates of the
reduced density matrix for the spin ($\rho_2$), or Schmidt states.
These states,
along with the eigenvalues, provide the most concise description of
the state of the spin, and they describe the correlations
with the apparatus as well.   The Schmidt states start out being
very different from the the $\{\ku ,\kd \}$ basis, but approach
very close as the measurement is completed.  The Schmidt states
then stabilize close to the $\{\ku ,\kd \}$ states and do not
change much after the measurement.

The degree to which the Schmidt states are $\{\ku ,\kd \}$ can be
studied by examining the off diagonal matrix element of $\rho_2$ in
the $\{\ku ,\kd \}$ basis:
\begin{equation}
\bu \rho_2 (t)\kd = a b^{\ast} \langle X_{\downarrow}(t)|
X_{\uparrow}(t)\rangle.
\label{offdiag}
\end{equation}
Equation \ref{offdiag} shows that to the extent that the overlap of
$X_{\uparrow}(t)\rangle_1$ with  $|X_{\downarrow}(t)\rangle_1$
is small, $\{\ku ,\kd \}$ are the
eigenstates of $\rho_2$.  As discussed in
Appendix A, the typical overlap goes down as the
size of the apparatus is increased, so even
in this simple example one can see that large
size is an advantage when building an apparatus.

\subsection{A Catch}
\label{ac}

The case where $a$ and $b$ are nearly equal deserves special attention.
In this case the eigenvalues of $\rho_2$ become nearly degenerate
at late times, and the form of the eigenstates of $\rho_2$
becomes a delicate matter.

Consider  first the case of strict equality:
\begin{equation}
a = b = 1/2
\label{aeqb}
\end{equation}
In this limit
$\bu \rho_2 (t)\ku =  \bd \rho_2 (t)\kd = 1/2$, and the form of
the eigenstates is completely determined by $\bu \rho_2(t) \kd$.
In this special case the eigenstates are either
\begin{equation}
\frac{(\ku
\pm e^{i\theta}\kd)}{\sqrt{2}},
\label{badeigenstates}
\end{equation}
if $\bu \rho_2(t) \kd$ is non-zero (no matter how small!), or
undetermined if $\bu \rho_2(t) \kd$ is {\em exactly} zero.
(See Appendix A
for the definition of the phase $\theta$ and further details.)

A physicist need never worry about the ``measure zero'' case where
Eq \ref{aeqb} is exactly obeyed, but there is a more general point to
be made:
As $a$ and $b$ get close to one another, even very small
values of $\bu \rho_2(t) \kd$ can be ``too large''
and cause the eigenstates of $\rho_2$ to deviate greatly
from the desired $\ku$ and $\kd$ states.

In the toy system, the mean magnitude  of $\bu \rho_2(t) \kd$ is never
zero, although it can be made arbitrarily small by increasing
the size of the apparatus.  Thus for every apparatus size
there exists a limit to how close $a$ and $b$ can get without
causing the Schmidt states to exhibit large deviations from the desired
behavior.
On the other hand, given any arbitrarily close values of
$a$ and $b$, there exists a sufficiently large choice of
$n_1$ so that the desired behavior is achieved. In Appendix A I show that
the minimum value of $|a - b|$ scales as $1/\sqrt{n_1}$. (Note that
the ``$n_1$'' of a real macroscopic apparatus is huge!)

I also argue in Appendix A that if one accepts the departure of
$\bu \rho_2(t) \kd$ from zero as an indication of the precision of
the apparatus, then there is nothing particularly wrong with the apparatus
in the $a \rightarrow b$ limit. The apparatus is just unable to
precisely resolve the value of $a-b$.  One could even argue that the
sensitivity of the Schmidt basis to the precise value of $\bu
\rho_2(t) \kd$ in this limit makes the Schmdit basis misleading when
$|a-b|$ falls below the ``experimental resolution''.    (This amounts
to a major concession to W. Zurek,
with whom I have been having ongoing informal debates about the value of the
Schmidt decomposition!)

\subsection{Some Red Herrings}
\label{srh}

I came up against the special behavior discussed in Section
\ref{ac} early in the course of this work. Although I
appreciated the overall delicacy of the degenerate
eigenvalue case, my efforts to preserve the desired behavior
in that limit were not always to the point.   In this
subsection I critique some remarks on this subject in
previous papers of mine.

The eigenstates of $H_I$ (from Eq \ref{iham}) have the form of either:
\begin{equation}
|\lambda_I\rangle = \ku\otimes |\lambda_{\uparrow}\rangle
\label{lup}
\end{equation}
or
\begin{equation}
|\lambda_I\rangle = \kd\otimes |\lambda_{\downarrow}\rangle
\label{ldown}
\end{equation}
where the $|\lambda_{\uparrow}\rangle$ and $|\lambda_{\downarrow}\rangle$
are the eigenstates of $H^{\uparrow}_1$ and $H^{\downarrow}_1$
respectively.
The addition of sub-dominant ``self-Hamiltonians'' for the two subsystems
does not have a large overall effect.  However, frequently a handful of
energy eigenstates deviate greatly from Eqs \ref{lup} and \ref{ldown}.
(The reason is that among the random set of energy eigenvalues
there are always a few which are quite close together.
Under such circumstances small perturbations can greatly affect the
form of the eigenstates, as was already discussed regarding $\rho_2$.)

Some of my previous efforts to produce a ``good measurement'' in the
$a \rightarrow
b$ limit focused on avoiding the bad energy eigenstates, which are
not close to Eqs \ref{lup} and \ref{ldown}.  In \cite{a92q}
(section 6.1) I further reduced the coefficients of the self-Hamiltonians,
and in \cite{a91t} I specially chose the initial conditions to avoid
the bad energy eigenstates.  In fact none of these efforts were
useful, because they did not reduce the overlap of
$|X_{\uparrow}(t)\rangle_1$ with $|X_{\downarrow}(t)\rangle_1$.
This overlap is present even when the energy eigenstates are exactly
given by Eqs \ref{lup} and \ref{ldown}, and the form of the eigenstates
was not the problem which needed addressing.  The overlap is most
easily reduced in the toy system by increasing the the size ($n_1$)
of the apparatus.

\section{Consistent histories and the Hamiltonian}
\label{chath}

The states $\ku$ and $\kd$ for the spin are absolutely stable under
the action of $H_I$.  Once one projects with $\hat{P}_{\uparrow}$
the projected state will remain of the form $\ku \otimes |
{\rm something}\rangle_1$ from then on.  Projecting with $\hat{P}_{\uparrow}$
and  $\hat{P}_{\downarrow}$ at different times is certain to
give zero. If $\{\ku, \kd\}$ is the projection basis, then
the only path projected states with non-zero amplitude have all the projections
either uniformly up or uniformly
down (regardless of the values and frequency of the projection
times).  The only cross-term in Eq \ref{sumid} which could potentially cause
sum rule violation is the dot product between the uniformly
up and uniformly down path projected states.  This cross-term
is also zero because $\bu \downarrow\rangle = 0$.
Thus using the $\{\ku, \kd\}$ projection basis is sure to
give consistent histories for this model, no matter what the initial
state.  Although these histories do not explicitly exhibit dynamics
associated with the evolving correlations, their consistency is
closely linked with these dynamics via the special stability of the
$\{\ku, \kd\}$ basis.  In Zurek's \cite{z82} language, the $\{\ku,
\kd\}$ basis is a ``pointer basis'', which does not loose quantum
coherence via the interactions.

The $\{\ku, \kd\}$  basis is special because of the form of
$H_I$ (Eq \ref{iham}).  The basis $\{|{\cal I}\rangle,
|\perp\rangle\}$ is nothing special from the point of view of
the Hamiltonian, and it is not surprising that consistent
histories were not found using that projection basis.

The other sets of consistent histories discussed in Section 9
involved projecting first on the Schmidt states and then on the
$\{\ku, \kd\}$  basis.  After the first projection,  the
two resulting path projected states are just the two terms
($\sqrt{p_1}|1\rangle^S_2 \otimes
|1\rangle^S_1 $ and $\sqrt{p_2}|2\rangle^S_2 \otimes
|2\rangle^S_1$) of the Schmidt decomposition of the total
wavefunction at $t_1$.  The subsequent evolution of each path
projected state may be treated as in Eq \ref{fcex}:
\begin{eqnarray}
|[1](t)\rangle = e^{-i(t - t_1)H_I}\left(\sqrt{p_1}|1\rangle^S_2 \otimes
|1\rangle^S_1 \right) &=& \ku \left(\sqrt{p_1}\bu 1\rangle^S_2\right)
\otimes |1_{\uparrow}(t)\rangle_1 \nonumber \\
& & + \kd \left(\sqrt{p_1}\bd 1\rangle^S_2\right) \otimes
|1_{\downarrow}(t)\rangle_1
\label{p1ex}
\end{eqnarray}
where $ |1_{\uparrow}(t)\rangle_1 $ and $  |1_{\downarrow}(t)\rangle_1
$ are $ |1\rangle_1^S$ evolved under
$H^{\uparrow}_1$ and $H^{\downarrow}_1$ respectively.
Likewise:
\begin{eqnarray}
|[2](t)\rangle = e^{-i(t - t_1)H_I}\left(\sqrt{p_2}|2\rangle^S_2 \otimes
|2\rangle^S_1 \right) &=& \ku \left(\sqrt{p_2}\bu 2\rangle^S_2\right)
\otimes |2_{\uparrow}(t)\rangle_1 \nonumber \\
& & + \kd \left(\sqrt{p_2}\bd 2\rangle^S_2\right) \otimes
|2_{\downarrow}(t)\rangle_1.
\label{p2ex}
\end{eqnarray}
Given the form of Eqs \ref{p1ex} and \ref{p2ex} it is easy to see
the effect of later projecting on the $\{ \ku , \kd\}$ basis.
The resulting four path projected states are just the two terms
from  Eq \ref{p1ex} and the two terms from  \ref{p2ex}.
The relevant cross-terms, which must be zero to give
consistent histories are:
\begin{equation}
\langle [\uparrow,1]|[\uparrow,2]\rangle \propto \langle
1_{\uparrow}(t) | 2_{\uparrow}(t)\rangle
\label{ode1}
\end{equation}
and
\begin{equation}
\langle [\downarrow,1]|[\downarrow,2]\rangle \propto \langle
1_{\downarrow}(t) | 2_{\downarrow}(t)\rangle
\label{ode2}
\end{equation}
The quantity  in Eq \ref{ode1} is exactly zero because
$|1_{\uparrow}(t)\rangle$ and $|2_{\uparrow}(t)\rangle$
started orthogonal, and were unitarily evolved by the same Hamiltonian,
so they must remain orthogonal.  Likewise for Eq \ref{ode2}.

Unlike the first set of consistent histories discussed,
these histories are consistent because of orthogonality of
the path projected states in the {\em apparatus} subspace.
One can say that records of the spin at $t_1$ are present
in the apparatus.   The Schmidt decomposition (at $t_1$) was use to
resolve these records. (If any other projection basis had been
used at $t_1$, the counterparts of $|1_{\uparrow}(t)\rangle$
and $|2_{\uparrow}(t)\rangle$ would {\em not} have
started out orthogonal, and the cross-terms would not have come
out zero.)  The choice of second projection was also crucial.
By choosing the stable $\{ \ku, \kd\}$ basis, one avoided
loosing track of the records between $t_1$ and $t_2$, when the second
projections were made.

\section{Comparing consistent histories with instantaneous probabilities}
\label{cchwsp}

The consistent histories approach involves  assigning probabilities
to histories.  In contrast,
the wavefunction at
a  particular time can be used to assign a probability to any {\em state}
(possibly a state specified only for a subsystem).  One just projects
onto the state in question and squares to get the
probability.
  This
procedure can be repeated at different times (always evolving the
whole unprojected wavefunction). The
Schmidt paths just give a way of following the probabilities assigned
to a particular set of states.  Often these Schmidt states are very
interesting because they exactly reflect the correlations which are
present.

The consistent histories formalism actually coincides with the
instantaneous probabilities view in the special case where
the ``paths'' are defined at a single instant, utilizing just one
projection time. In this case
there is no difference between assigning a probability to a ``path''
or a state.  In fact, such probabilities automatically obey all
necessary sum rules, which is why no additional consistency conditions
are discussed when taking the instantaneous probabilities view.

The consistent histories formalism allows one to go beyond
the instantaneous view and (at the cost of extra conditions)
assign probabilities to { \em extended} histories.  That
is,  histories defined over more than  one
moment in time.  A number of authors
(for example
\cite{jh90,g-m&h92,d&h92})
have attached great importance to this way of going beyond
the instantaneous point of view.

In general the extended  histories and the instantaneous
probabilities offer very different points of view.
However, the the two
can be quite similar in the particular
case where
good measurements are made within
the closed system being treated.  In that case projecting on a
particular set of records (at a single time)
should be completely equivalent to projecting (even at  {\em
earlier} times) on the
corresponding state of the system being recorded, as long as the
projection is  made at a time {\em after} the measurement has been
completed (see for example refs \cite{g-m&h91,jh90} ).
Since the Schmidt decomposition can be applied to expose the
correlations among all the relevant
apparatuses and systems, one might expect that the paths traced out by
the Schmidt
states {\em after} a measurement should bear a lot of resemblance to one
set of
(extended) consistent
histories.  However, based on the work in this paper, it does not seem
that the extended consistent
histories and the Schmidt paths bear much
resemblance {\em during} the measurement process, when the correlations
are actually being set up.

Much of this paper
studies
the measurement process from the instantaneous probabilities
point of view.  It has rightly been pointed out
\cite{jh90,g-m&h90,d&h92,g-m&h92}
that the information provided by a wavefunction at a single moment in
time is of limited use in investigating many important issues in
quantum mechanics.  None the less, by following the {\em time
development} of the instantaneous probabilities
one is
able to provide some useful insights into the nature of the quantum
measurement.  (This is the used  in the pioneeing work by Zeh and Zurek.
The whole notion of Zurek's ``pointer
basis''\cite{z82} or Zeh's ``stability of the Schmidt
states''\cite{z73} is connected with
time evolution, as is the issue of ``permanence'' of the record, which
both these authors address.  Their analysis, which involves the time
development of instantaneous probabilities, is very different from
just looking at a wavefunction at a single moment in time.)

Like Zeh and Zurek, I have found the time development of the reduced
density matrix to offer a convenient perspective on the measurement
process.  One can answer
questions such as ``how long does the measurement take?''\cite{wz86,p&h&z93},
and ``what
is the state of the system half-way through the measurement?''(Fig 2).  In
turn these insights can help one deduce the features which make a good
measurement apparatus.

In this particular application I have not found the generalization to extended
histories offered by the consistent histories formalism
particularly illuminating.  No single set of extended
histories appeared to be following the correlations in any continuous
way, and no set indicated the
duration of the measurement process.  No doubt these features  can
be extracted by
considering a large number of {\em different} sets of extended consistent
histories but not in a particularly
direct way.  In short,
the time development of the reduced density matrix seems to allow
one to focus more directly on the measurement process, as compared
with the extended  histories point of view.

This is  not to say that the different focus offered by the
extended  histories is ``bad''.   After
all, in many realistic situations one does not {\em want} to focus on
the details of the measurement process. For example, the
``constant'' spin up and down consistent histories are probably
exactly how an observer who measures the spin in the $\{\ku ,\kd\}$
would want to think of the history of the spin.  Whether the spin was
once in a coherent superposition of up and down, and whether some
other system had {\em already} measured the spin in  the  $\{\ku
,\kd\}$ basis before his
measurement occurred would be of no practical interest to the
observer\footnote{For some related ideas see \cite{z92t}}.

\section{The arrow of time}
\label{taot}

As has been noted, for example by Zurek \cite{z82} and Zeh
\cite{z71,z90,zbook}, there
is an arrow of time built into the
dynamics discussed here.
This is dramatized in Fig 3, which is identical to Fig 1, but with the
time axis extended to the interval $[-2,2]$.  One can seen that the
pure ``initial''
($t=0$) state (which has zero entropy for the spin), is a very special state
and the ``collapse of the wavefunction'' proceeds in the direction of
increasing spin entropy.  The $t<0$ part of Fig 3 illustrates an
``un-collapsing'' wavefunction, where the correlations present
between spin and apparatus at early times are lost, and the pure
state emerges at $t=0$.  Then, for positive values of $t$ correlations
are established again.  The stability of these correlations (and thus
the goodness of the measurement) depend on another  such ``entropy dip''
not occurring for $t>0$.  In the language of Section 7,
one is depending on the random evolution of
$|X_{\uparrow}(t)\rangle_1$ and $|X_{\downarrow}(t)\rangle_1$ not to
cause these two states to overlap appreciably at later times.
(This issue has been discussed at length in \cite{z82}.)
Even the simple system discussed here is complex
enough for such large entropy dips to occur very rarely.  Still, with such a
small  apparatus, noticeable fluctuations are present.
(Note that the portion of Fig. 3 which is shown in Fig 1 is
uncharacteristically well behaved.  See Appendix C for further discussion.)

Aside from questions of stability, how fundamentally is the arrow of
time linked to quantum measurement?  The initial state,
$|\psi_i\rangle$  has zero entropy for the spin, so it is not
surprising that just about anything will cause the entropy to
increase.  What about starting with a more general initial state?
Schmidt tells us that (in a suitable basis) the most general state can
be written
\begin{equation}
|\tilde{\psi}_i\rangle = \sqrt{p_1}|1\rangle_2\otimes|1\rangle_1
+ \sqrt{p_2}|2\rangle_2\otimes|2\rangle_1.
\label{gic}
\end{equation}
I show in Appendix B that if one requires evolution which
generalizes Eq (\ref{fc}) to give
\begin{eqnarray}
|\tilde{\psi}_i\rangle & \rightarrow & |\tilde{\psi}_f\rangle \nonumber \\
& = & \sqrt{p_1}\left(\langle \uparrow | 1\rangle_2 \ku\otimes|A\rangle_1
+ \langle \downarrow | 1 \rangle_2 \kd\otimes |B\rangle_1 \right)
\nonumber
\\
& & + \sqrt{p_2}\left(\langle \uparrow | 2\rangle_2 \ku \otimes|C\rangle_1
+ \langle \downarrow | 2 \rangle_2 \kd \otimes |D\rangle_1 \right)
\label{gfc}
\end{eqnarray}
then one must have increasing (or constant) entropy of the spin
($-\tr[\rho_2\ln(\rho_2)]$) as $|\tilde{\psi}_i\rangle  \rightarrow
|\tilde{\psi}_f\rangle$.  Thus ``good measurement'' appears to be
closely linked with increasing entropy, even for high entropy initial
states.  (Note that I have chosen all four
apparatus states,  $|A\rangle_1$, $|B\rangle_1$, $|C\rangle_1$, and
$|D\rangle_1$ to be mutually orthogonal.  This means that in
$|\tilde{\psi}_f\rangle$  the
apparatus has a record of whether the spin is up or down,
{\it and} which term of Eq (\ref{gic}) has been ``measured''.)

\section{Conclusions}
\label{c}

The ideas put forward by Zeh \cite{z73}, Zurek \cite{z82}, Joos and
Zeh \cite{j&z85}, and
Unruh and Zurek \cite{u&z89}, have sufficiently de-mystified the
notion of wavefunction collapse that one can actually unitarily
follow the evolution of a system right through the collapse process.
I have investigated a simple system which exhibits ``wavefunction
collapse''.  I find Zeh's idea of
watching the evolution of the eigenstates of the reduced density
matrix (Schmidt paths)  particularly appealing.  This approach allows
one to follow
exactly the evolution of the correlations among subsystems.  It also
allows one to visualize the collapse process quite explicitly, as
illustrated in Fig 2.
However, when eigenvalues of the reduced density matrix are nearly
degenerate, the eigenstates become very sensitive to ``noise'', and
can give a misleadingly unstable picture of what is going on.

I also applied the ``consistent histories'' analysis (of Griffiths
\cite{g84}, Omnes \cite{o88a,o88b,o88c} and Gell-Mann and Hartle
\cite{g-m&h90}) to the same
system.  In one limit, this approach can reproduce the reduced density
matrix results where probabilities are assigned at instants in time.
More generally, the consistent histories allow  one (when the consistency
conditions are
satisfied) to assign probabilities to extended
histories of the system.
In the example studied here, many different
sets of histories passed
the consistency test.  It is intriguing that one set of consistent
histories for the spin did not reflect the interesting evolution of
the correlations between the spin and the apparatus.
Instead, it was more a reflection of the stability
properties of the spin. That set of
histories would look the same for a static spin, decoupled from the
apparatus.  Other consistent histories exhibited more direct links
to the ``quantum measurement'' process underway.  However, there was
very little resemblance between any given set of extended consistent histories
and the Schmidt paths for the system.  I argued that the reduced
density matrix offered a more convenient point of view from which to
analyze the measurement process.

I have employed a perspective on wavefunction collapse which
explicitly does {\em not} make a choice among the possible outcomes at
the end of the
measurement process.  This results in Everett's ``many
worlds''.
An advantage of this perspective is that the question of what makes a
good apparatus can be addressed quite directly.  To this end,
I have discussed in detail the Hamiltonian used to evolve
the system, and the features necessary to accomplish
the desired evolution.  Even in this primitive example, the
quality of the apparatus is very clearly linked with its size, and with the
statistical ``irreversibility'' associated with the thermodynamic
arrow of time.
33
\section{Acknowledgments}

I would like to thank W. Zurek,  M. Joyce, J. Halliwell and A.C.
Albrecht for some
very helpful discussions.
This work was supported in part by the DOE and
the NASA (grant NAGW-2381) at Fermilab.

\vfill
\pagebreak
\appendix
\section{Nearly degenerate density matrices}
\label{nddm}

\subsection{Mathematics}

Consider the matrix:
\begin{equation}
\left(
\begin{array}{cc}
1/2 + \delta & \omega^{\ast} \\
\omega & 1/2 - \delta
\end{array}
\right).
\label{matrix}
\end{equation}
Its (un-normalized) eigenstates are:
\begin{equation}
\left\{
\begin{array}{c}
\left(
\begin{array}{clc}
\frac{\delta - \sqrt{\delta^2 + \omega^{\ast}\omega}}{\omega}&,&1
\end{array}
\right)\\
\left(
\begin{array}{clc}
\frac{\delta + \sqrt{\delta^2 +
\omega^{\ast}\omega}}{\omega}&,&1
\end{array}
\right)\\
\end{array}
\right\}.
\label{eigenstates}
\end{equation}
If one takes the limit $\omega \rightarrow 0$ while keeping
$\delta$ fixed the eigenstates become proportional to
$(0,1)$ and $(1,0)$.  This is what the toy model is trying
to accomplish for $\rho_2$ at late times, by making the
off-diagonal terms (here $\omega$) small.
However, if one takes $\delta \rightarrow 0 $ while
keeping $\omega$ fixed the eigenstates become proportional
to $(\pm\frac{
|\omega|}{\omega},1)$. (Since $\omega$ corresponds to
$\bu \rho_2 \kd$, $\theta$ in Eq \ref{badeigenstates}
is just the phase of $a b^{\ast} \langle X_{\downarrow}(t)|
X_{\uparrow}(t)\rangle$.)

If one wants to require the eigenstates to be close to $(0,1)$
and $(1,0)$, one can require:
\begin{equation}
\frac{\delta - \sqrt{\delta^2 + \omega^{\ast}\omega}}{\omega} < \epsilon
\label{smallterm}
\end{equation}
for some small epsilon.  For small values
of $|\omega|/|\delta|$ Eq \ref{smallterm} becomes
\begin{equation}
\delta > \frac{\omega}{2\epsilon}.
\label{smallterm2}
\end{equation}
This paper involves the case where $\omega$ is the overlap
of two random vectors in a space of size $n_1$.  The
magnitude of such a quantity is the net distance
traversed by an $n_1$-step random walk with average step size
proportional to $\sqrt{n_1}$, so $\omega \propto 1/\sqrt{n_1}$.
Combining this with Eq \ref{smallterm2}, on can see that
the minimum allowed value of $\delta$ goes as $1/\sqrt{n_1}$.

\subsection{Physics}

The goal of the interactions was to get the wave function into the
form given by Eq \ref{fc}:
\begin{equation}
|\psi_f\rangle = a\ku_2\otimes|Y\rangle_1 + b\kd_2\otimes |Z\rangle_1
\label{fc2}
\end{equation}
with
\begin{equation}
\langle Y|Z\rangle = 0.
\label{aorth}
\end{equation}
  When $a\rightarrow b$ the Schmidt
decomposition tells us that if one insists on Eq \ref{aorth} holding
exactly
(which is what Schmidt does), then the Schmidt expansion can look very
different than Eq \ref{fc2}.  This is because even small
non-zero values of $\langle
X_{\uparrow}(0)|X_{\downarrow}(0)\rangle $ ($\equiv \omega$ in Eq
\ref{matrix}) can have a large impact on the eigenstates in this
limit.  However, how {\em badly} does the wavefunction deviate from
Eq \ref{fc2} when $|\langle
X_{\uparrow}(0)|X_{\downarrow}(0)\rangle| >> |a-b|$ (that is, when $1>>\omega
>> \delta$ in Eq \ref{matrix})? One check is to look at the
``overlap'' between actual reduced density matrix for the spin
($\rho_2$ given by Eq \ref{matrix}, for example) and the ideal result
\begin{equation}
\rho_I \equiv \left(
\begin{array}{cc}
1/2 + \delta & 0 \\
0 & 1/2 - \delta
\end{array}
\right).
\label{matrixI}
\end{equation}
The quantity $\tr\sqrt{\rho_2\rho_I}$ is a good measure of the
overlap which can be understood by writing each $\rho $ in terms of
its eigenstates\footnote{The square root of the operator
$\rho_2\rho_I$ is defined in the
usual way.  The operator is expressed in its eigenbasis and the square
root of its eigenvalues are taken.}.  The value of
$\tr\sqrt{\rho_2\rho_I}$ is unity when $\rho_2 = \rho_I$, and is zero when
no eigenstates (with non-zero eigenvalues) overlap.
Taking $\rho_2$ from Eq \ref{matrix} and $ \rho_I$ from Eq
\ref{matrixI}, and expanding for small $\omega$ (keeping $\delta $
fixed) one gets
\begin{equation}
\tr\sqrt{\rho_2\rho_1} = 1+O(w^{2}).
\label{answer}
\end{equation}
At least according to this measure, $\rho_2$ and $\rho_I$ are very
close when $\omega$ is small, even when their eigenstates are very different.

	One can further explore the suitability of the system as a
``measurement apparatus'' in the $a \rightarrow b$ limit by
considering the interaction of a third system with the apparatus.
One feature of a good apparatus is a ``pointer'',  which clearly
exhibits the outcome of the measurement, and which can subsequently be
measured by other systems to determine the outcome of the original
measurement. (Zurek \cite{z82}  emphasizes this point by clearly
partitioning out the pointer from the rest of the apparatus).  As
discussed earlier, this feature is absent from the toy model.

For the sake of discussion I will force the issue by assuming there is
a third system which has detailed information about
$|X_{\uparrow}(t)\rangle$.  It can use this information to suitably
measure the apparatus.  If  the apparatus is found in
$|X_{\uparrow}(t)\rangle$ the third system will conclude that the spin
is up.   The issue is being ``forced'' only in the sense that one is
asking the third system to know something very complicated (namely
$|X_{\uparrow}(t)\rangle$, a complete ``microscopic'' state with messy
time evolution) in order to use the apparatus.  In a good apparatus
some simple feature (such as a blip on a screen) should indicate the outcome.

So how much of a mistake does the third system make by   using this
procedure?  The errors come  because the overlap of $|\psi_f\rangle$
with $|X_{\uparrow}(t)\rangle$ receives contributions not just from
the first term  in Eq \ref{fcex}, which is indeed correlated with
$\ku$, but also from the {\em second} term (correlated with $\kd$) due
to the non-zero value of $\langle
X_{\downarrow}(t)|X_{\uparrow}(t)\rangle$.  To the extent that
$\langle X_{\downarrow}(t)|X_{\uparrow}(t)\rangle$ is small, the
errors are small {\em even} in the $a \rightarrow b$ limit.  The size of
$\langle X_{\downarrow}(t)|X_{\uparrow}(t)\rangle$ simply represents
the precision of the apparatus.

Repeated measurements by the third system of identically prepared
spin-apparatus systems should yield inferences of the values of $a$
and $b$.  These inferences should be increasingly good as the number of
repetitions increases.  The one ``problem'' encountered as
$a\rightarrow b$ is that the actual value of $a-b$ \linebreak falls below the
precision of the apparatus.  In this case the third system could only
conclude that $a\approx b$ within the experimental uncertainties. As
long as the precision of the apparatus is acceptable,  there is {\em
no} problem with the apparatus in the $a \rightarrow b $ limit.

However, by choosing to look at the Schmidt decomposition, one is
looking at something which can be very sensitive to $a-b$, as
illustrated at the beginning of this appendix.  In the case where the
magnitude of $a-b$ falls below the acceptable resolution of the
apparatus, one could argue that the Schmidt decomposition can be very
misleading.
For example, the spin-apparatus system could be in a state
sufficiently close to Eq \ref{fc} for practical purposes, but the
Schmidt decomposition could yield something that looks completely
different.

This more wary attitude toward the Schmidt decomposition represents a
step back from the enthusiasm I have expressed on other occasions (see
\cite{a92q} at the very end of section 2.2, for example).

\section{Generalized measurements and the arrow of time}
\label{gmatat}

The point of this Appendix is to show that the final state in
Eq \ref{gfc} has higher entropy (relative to the spin-apparatus
partition) than the generalized initial state given by Eq \ref{gic}.

Equation \ref{gic} is manifestly in Schmidt form, and
Eq \ref{gfc} can be put in Schmidt form by collecting
even and odd terms together.
For the initial state, the eigenvalues of the density
matrix are $p_1$ and $p_2$.  For the final state, the eigenvalues are
\begin{eqnarray}
p_{\uparrow} & = & p_1|\langle \uparrow|1\rangle_2|^2 \; +
\;  p_2|\langle \uparrow|2\rangle_2|^2
\label{pup}\\
p_{\downarrow} & = & 1 - p_{\uparrow}.
\label{pdown}
\end{eqnarray}
The fact the $ _1\!\langle A|C\rangle_1 =
 _1\!\langle B|D\rangle_1 = 0 $ is important for obtaining Eq \ref{pup}.

Since the entropy is monotonically decreasing in $|p_1 - p_2|$,
it will suffice to show that $|p_{\uparrow} - p_{\downarrow}|
\leq |p_1 - p_2|$.
Without loss of generality I take $p_{\uparrow} >
p_{\downarrow}$ and $ p_1 > p_2 $.
\begin{equation}
p_{\uparrow} - p_{\downarrow} = 2p_\uparrow - 1 =
2\left( p_1|\langle \uparrow | 1 \rangle _2|^2 \; + \;
p_2 | \langle \uparrow | 2 \rangle_2 |^2 \right) - 1.
\label{step1}
\end{equation}
Now define: $\Delta \equiv p_1 - p_2$.  Using this
definition and $p_1 + p_2 = 1$ on can rewrite Eq \ref{step1} as:
\begin{eqnarray}
p_{\uparrow} - p_{\downarrow} & = &
2\left( \frac{1+\Delta}{2}|\langle \uparrow | 1
\rangle _2|^2 \; + \; \frac{1-\Delta}{2} | \langle
\uparrow | 2 \rangle_2 |^2 \right) - 1 \label{step2} \\
& = & \Delta \left(
| \langle \uparrow | 1 \rangle _2|^2 \; -\;
| \langle \uparrow | 2 \rangle_2 |^2
\right) \label{step3} \\
& = & \Delta\left( 2 |\langle \uparrow | 1 \rangle _2|^2
- 1 \right) \label{step4}
\end{eqnarray}
where the normalization condition  $\langle \uparrow |
\uparrow \rangle = | \langle \uparrow | 1 \rangle _2|^2 \; +\;
| \langle \uparrow | 2 \rangle_2 |^2 =
 1 $ was used in the final step.  Since
$\left( 2 |\langle \uparrow | 1 \rangle _2|^2 - 1\right)$
is manifestly bounded above by unity, the desired result,
$|p_{\uparrow} - p_{\downarrow}|  \leq |p_1 - p_2|$, is obtained.

\section{Search technique}
Figure 1 is a ``blow up'' of a small portion of Fig 3.  The reader
might have noticed that the portion shown in Fig 1 is much closer
 to
the ``desired behavior'' than any other portion of Fig 3.  This is
due to the fact that I did a fair amount of fiddling around, trying to
choose parameters which would make a good quantum measurement.  The
time range I looked at while searching parameter space was the same
range used in Fig 1.  Given this search ``technique'', it is not
surprising that my search ended on an atypical case.  I stopped when I
had found what I wanted (within the window of Fig 1).  On could
say that Fig 1 is slightly misleading.  On the other hand, one could
just as well say that I understand the apparatus:  I am able to prepare
it in a suitable manner so that a good measurement is performed, and
the record is kept for a specified period (in this case, .2 units of
time).
Just about any apparatus must be dealt with in this way.

\pagebreak
\section*{Figure Captions}

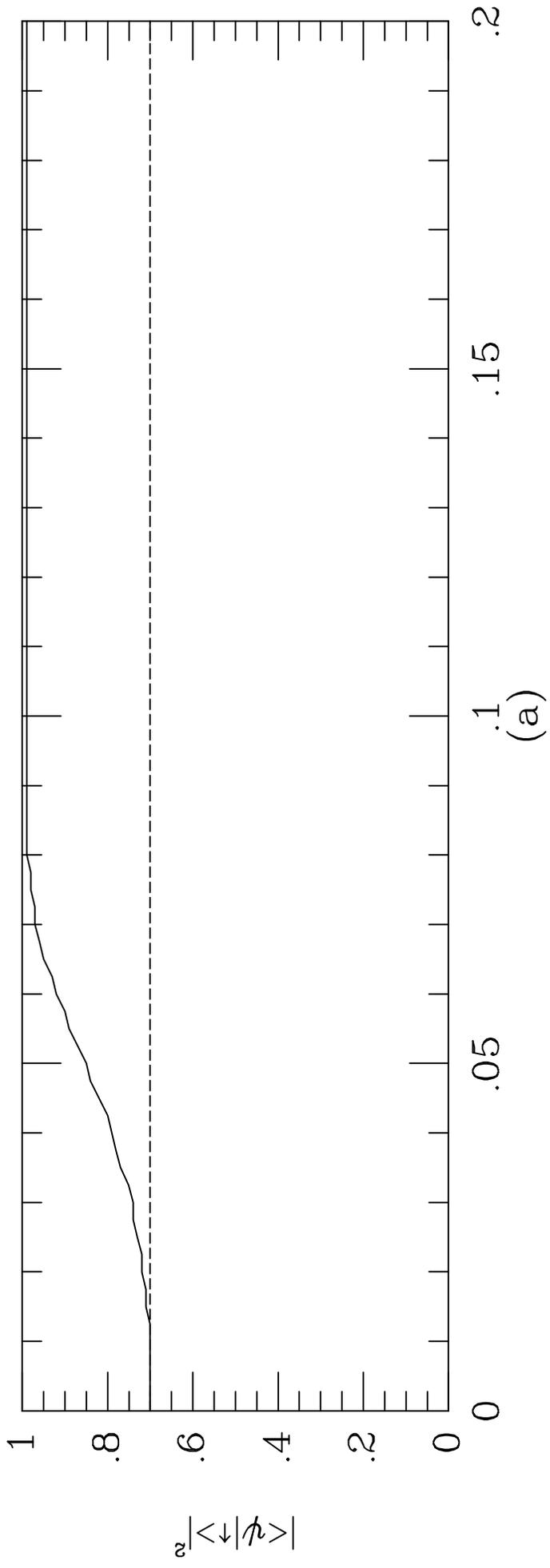
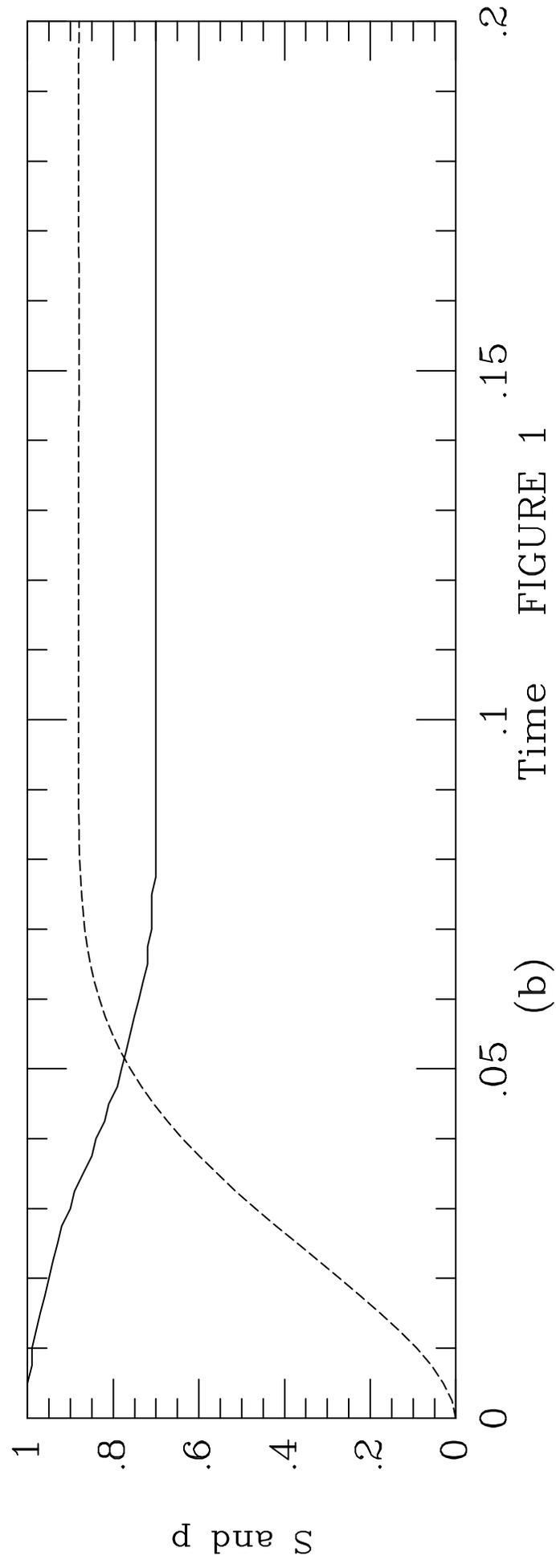
Figure 1:{
a: The solid curve is
$|\langle \uparrow|1\rangle^S|^2$, and the dashed curve gives  $\bu\rho_2\ku$.
b: The solid curve is the largest eigenvalue of $\rho_2$,
the dashed curve is the entropy of the spin. }

\vspace{3.3in}
Figure 2:{ ``A collapsing wavefunction.'' Each plot depicts an
eigenstate of $\rho_2$ in terms of  $p({\uparrow}) \equiv  |\langle
\uparrow|
i\rangle|^2$ and $p({\downarrow}) \equiv |\langle \downarrow |
i\rangle|^2$.  The columns correspond to three different times. The
two rows correspond to the two eigenstates.
}

\vspace{3.3in}
Figure 3: {The same plots as Fig 1 extended over a wider time  range.}

\vfill

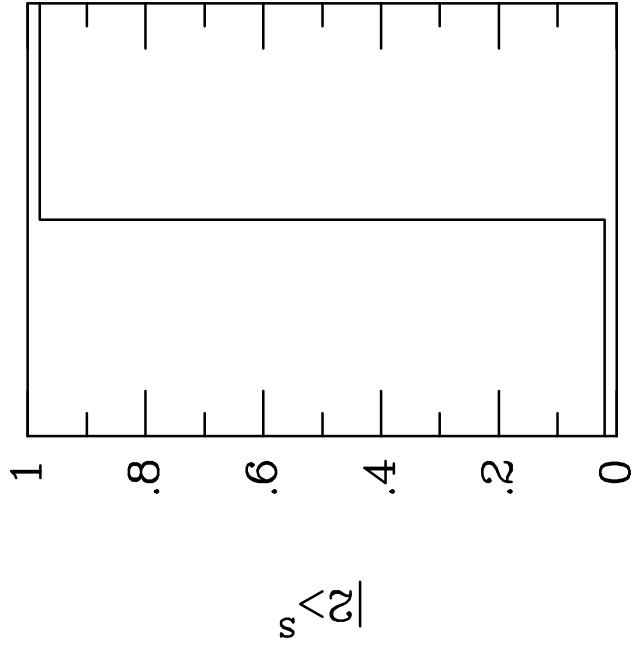
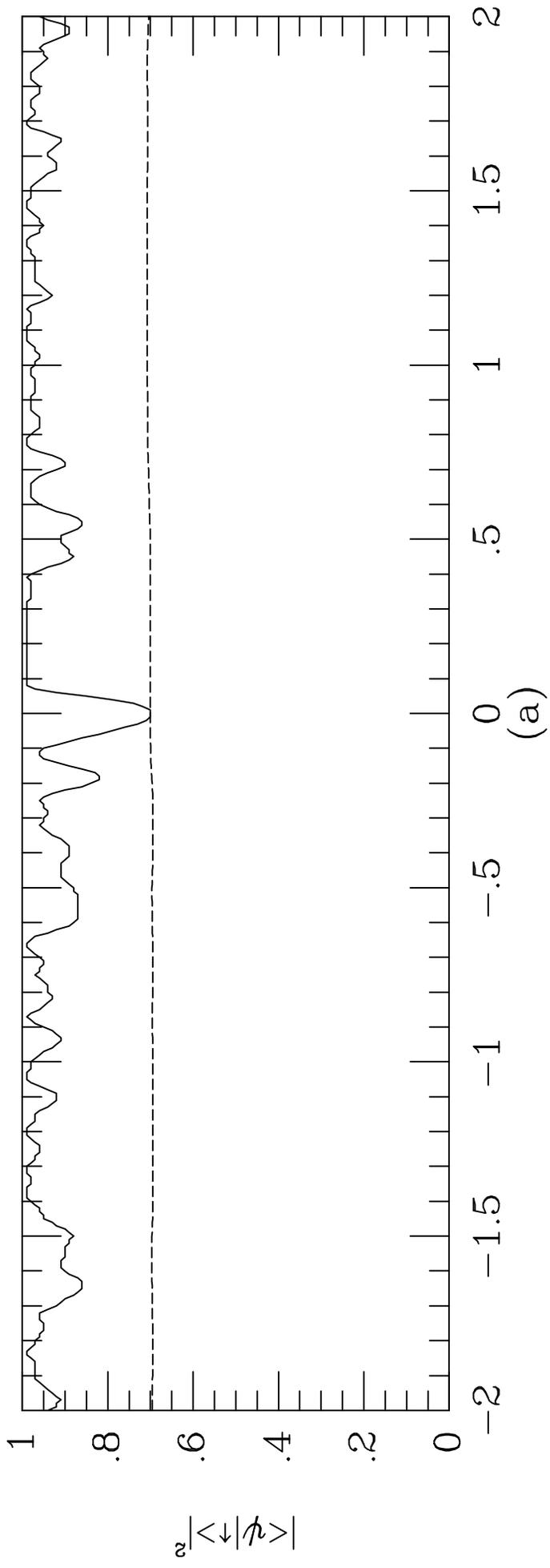
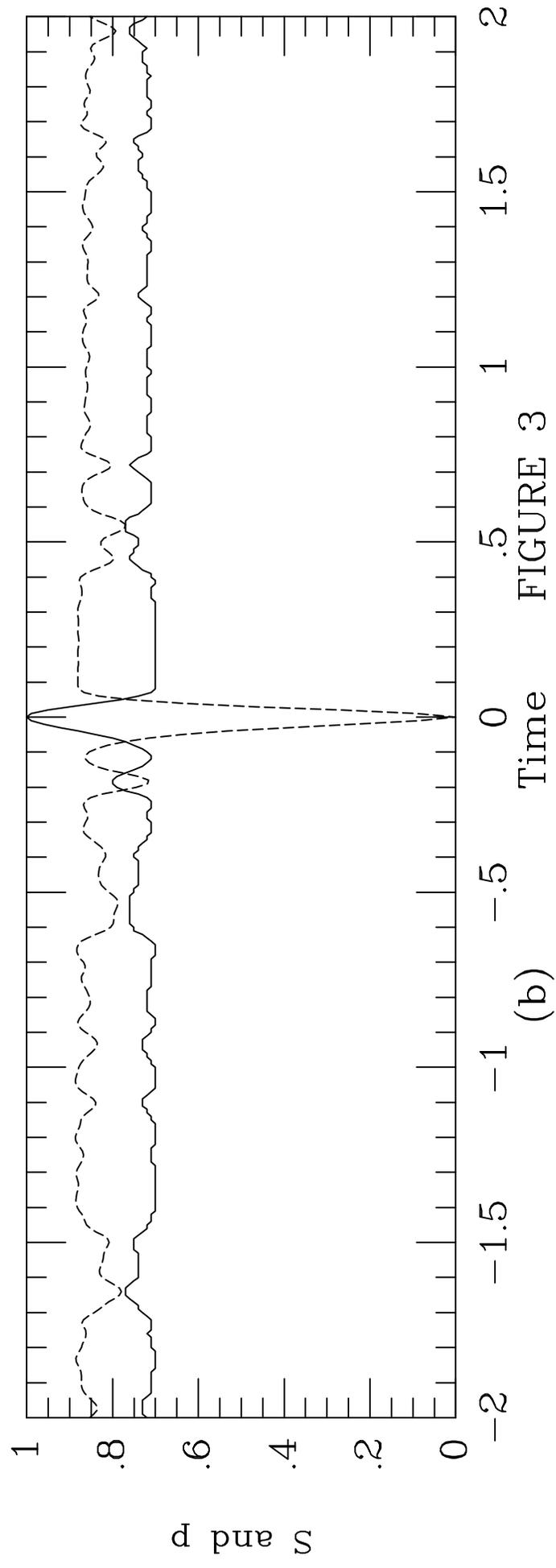
\end{document}